\documentstyle[12pt]{article}
\begin{document}
\newcommand{\be}{\begin{equation}}
\newcommand{\ee}{\end{equation}}
\title{A technique for generating Feynman diagrams}
\author{Sigurd Schelstraete \thanks{Research assistant NFWO,
e-mail: schelstr@allserv.rug.ac.be} \\
Henri Verschelde \thanks{Senior research associate NFWO,
e-mail: hschelde@allserv.rug.ac.be} \\
Department of mathematical physics and astronomy \\
University of Ghent \\
9000 Ghent \\
Belgium}
\maketitle
\begin{abstract}
We present a simple technique that allows to generate Feynman diagrams for
vector models with interactions of order $2n$ and similar models
(Gross-Neveu, Thirring model) using a
bootstrap equation that uses only the free field value of the energy
as an input. The method allows to find the diagrams to, in principle,
arbitrarily high order and applies to both energy and correlation functions.
It automatically generates the correct symmetry
factor (as a function of the number of components of the field)
and the correct sign for any diagram in the case of fermion
loops. We briefly discuss the possibility of treating QED as a
Thirring model with non-local interaction.
\end{abstract}
\newpage
\setlength{\unitlength}{1mm}

\section{Introduction}
In field theory the calculation of physical quantities is commonly
given as a sum of Feynman diagrams. Each diagram is calculated with
a certain weight, dependent on the number of symmetries of the
diagram. A rule to find this symmetry factor is given in
\cite{wu,goldberg}. Furthermore expressions were found for the
total number of diagrams that contribute in any order
to various quantities of interest \cite{jacobs,cvitan}. In
order to find the number and type of diagrams that contribute
to a physical quantity, numerous programs were developed
\cite{kaneko,prog}.
In this article we want to present a technique
for generating Feynman diagrams
that is a generalisation of a similar technique previously used in the
study of the vector model with quartic interaction
in 0+0 dimensions \cite{art}.
The results obtained in this paper are valid for
any number of dimensions. We will find all diagrams that contribute
to the energy and the correlation functions
(to, in principle, arbitrarily high order) with the correct symmetry
factor. This factor can be a function of the number of components of
the field. The technique is essentially different from the ones
that are usually applied tot generate diagrams, since it does not
involve any combinatorics. It is a recursive method that requires
only the free field value of the energy as an input.

In \cite{art} it was observed that
a particular relation between
the partial derivatives of the energy of the vectormodel could be used to
obtain a perturbation series for the energy.
In the simple case of the 0+0 dimensional vector model it also allowed to
find explicit expressions for the coefficients of the large $N$ expansion
of the model. A general form for this coefficients could be found and
proven to all orders. Unfortunately, a straightforward application
of the method used in 0+0 dimensions
to higher dimensions is not possible. However, if we slightly
adapt the original model a similar technique can be used in higher
dimensions.
We will consider here the vector model with quartic self-interaction.
Instead of considering a model with constant mass, we will allow the
mass to depend on the coordinates (the mass is therefore
an external field). The model is defined by the partition function:
\begin{equation}
{\cal Z} = e^{-E} = \int {\cal D} \phi \exp -{1 \over 2} \int
dx \partial_\mu \phi
\partial_\mu \phi + m(x) \phi ^2(x) - g \int (\phi ^2)^2 \ ,
\end{equation}
\noindent
where $\phi$ is a N-component field.
If we now consider the first and second functional derivative of $E$
with respect to $m(x)$ we find:
$$
{{\partial E} \over {\partial m(x)}} = {1 \over 2} <\phi ^2(x)>
$$
\noindent
and:
\begin{equation}
{{\partial ^2 E} \over {\partial m(x)} ^2} =
-{1 \over 4} <(\phi ^2(x))^2>  + ({{\partial E} \over {\partial m(x)}}) ^2,
\end{equation}
\noindent
where $<...>$ denotes the expectation value of an operator. Integrating
the last equation with respect to $x$ we find:
\be
<\int (\phi ^2)^2> = 4 \int (({{\partial E} \over {\partial m(x)}}) ^2
- {{\partial ^2 E} \over {\partial m(x)} ^2})
\ee
We rewrite this equation in the form that will be of interest to us:
\be
\label{bootstrap}
{\partial E \over \partial g} = 4 \int (({{\partial E} \over
{\partial m(x)}}) ^2  - {{\partial ^2 E} \over {\partial m(x)} ^2})
\ee
This is the generalisation of the formula found in \cite{art}. It is
clear how this formula can be used to generate a perturbation series
for the vectormodel. If we start with the zeroth order approximation
of the energy (i.e.: the value of the energy when there is no interaction),
formula (\ref{bootstrap}) gives us the derivative of the energy
(with respect to the coupling constant) to the same
order. Upon integration, we therefore find the energy to first order. This
again
can be used in (\ref{bootstrap}) to find the energy to second order. By
continuing this way we generate a perturbation series solely from the
knowledge of the free theory and the fact that the energy obeys
(\ref{bootstrap}). Of course, since the perturbation series is known
to be given by a sum of Feynman-diagrams
it is not unexpected to find that we will simply
generate the diagrams to any desired order, with the exact symmetry
factors.

\section{Generating Feynman diagrams}
In order to be able to carry out the procedure described in the previous
section, we must know the value of the energy when there is no interaction.
Without any difficulty this is found to be:
\be
\label{free}
E(g=0) = {N \over 2} Tr \log(-\partial ^2 + m)
\ee
\noindent
(remember that $m$ is a field, rather than a constant)
If we denote by $\{|x>\}$ a complete
set of eigenstates of the position operator, then (\ref{free}) can be
written as:
\be
\label{free2}
E(g=0) = {N \over 2} \int dy <y|\log(1-{m \over \partial ^2})|y>
\ee
For later convenience we have substracted a quantity that is independent
of $m$. This will not change the result. We will also denote $- \partial ^{-2}$
as $G$. The expression (\ref{free2}) is defined by its series expansion,
and so is the functional derivative with respect to $m(x)$. If we consider
a general term in the series expansion, the derivative will work
successively on each of the m's that occur in the term. We find that
the derivative of such a general term is given by:
\be
\label{genterm}
{\partial \over \partial m(x)} <y|(m G)^n|y> = \sum_{j=0}^{n-1}
<y|(m G)^j|x><x|G (m G)^{n-1-j}|y>
\ee
And when we integrate with respect to y:
\be
{\partial \over \partial m(x)} \int dy <y|(m G)^n|y> = n <x|G (m G)^{n-1}|x>
\ee
We therefore find:
\be
\label{der1}
{\partial E \over \partial m(x)} = - {N \over 2} <x|G {1 \over 1+m G} |x>
\ee

We will also need higher derivatives of the energy. Let us therefore consider
the following derivative:
\be
{\partial \over \partial m(x)} <y| G {1 \over 1+ m G} |z> =
{\partial \over \partial m(x)} <y| G \sum_{n=0}^{\infty} (- m G)^n|z>
\ee
Again using (\ref{genterm}), we find that this is equal to:
\be
- \sum_{n=0}^{\infty} \sum_{j=0}^n <y| G (- m G)^j|x> <x| G (- m G)^{n-j}|z>
\ee
By rearranging the double sum we can write this in the form:
\be
\label{double}
-<y| G {1 \over 1+ m G} |x> <x| G {1 \over 1+ m G} |z>
\ee
This formula has a simple interpretation. Since
$<y| G {1 \over 1+ m G} |z>$ is nothing but the propagator of the
field between points y and z, taking the derivative with respect to $m(x)$
is seen to double this propagator by inserting an extra point $x$
between the two endpoints (apart from a change of sign). We can denote
this graphicly as:

\begin{picture}(100,20)
%afleidingsregel voor propagator
\put(10,10){${\partial \over \partial m(x)}( y \hspace{1 cm} z) =
- y \hspace{1.5 cm} z$}
\put(24,10){\line(1,0){7}}
\put(49,10){\line(1,0){12}}
\put(55,10){\circle*{1}}
\put(55,12){$x$}
\end{picture}

We will now illustrate the techique by generating the first orders
in the series expansion of the energy. Starting from the lowest order
approximation (i.e. the value of the energy without interaction), we find
from the formulas given above:
\be
\label{afl1}
{\partial E \over \partial m(x)} = {N \over 2} <x|G {1 \over 1+m G} |x> + {\cal
O} (g)
\ee
\be
\label{afl2}
{\partial ^2 E \over \partial m(x)^2} =- {N \over 2} <x|G {1 \over 1+m G} |x>
<x|G {1 \over 1+m G} |x>  + {\cal O}(g)
\ee
Using (\ref{afl1}) and (\ref{afl2}) in (\ref{bootstrap}) and
integrating with respect to g (taking
into account the correct integration constant), we find:
\be
E = {N \over 2} Tr \log(1 + m G) + N(N+2) g
\int dx <x|G {1 \over 1+m G} |x>^2 + {\cal O} (g^2)
\ee
As long as $m$ is a function of $x$ the integral can not be carried out.
Unfortunately, we can not put $m$ equal to a constant at this point of the
calculation, because we need the full expression to find the higher
order terms. It is clear however that the first order expression we
have found  corresponds to the diagram:

\begin{picture}(100,20)
%E tot op orde g
\put(10,10){$E = {N \over 2}$}
\put(28,12){\circle{8}}
\put(35,10){$+ N(N+2)\  g $}
\put(65,12){\circle{7}}
\put(72,12){\circle{7}}
\put(68.5,12){\circle*{1}}
\put(80,10){$+ {\cal O}(g^2)$}
\end{picture}

as can be seen by drawing the propagators that occur in the term or by
considering the limit of constant m, in which case the term reduces to
$ (\int {dp \over p^2 + m})^2$, which is indeed
the value of the depicted diagram. This procedure can be continued
straightforwardly to obtain higher order diagrams.
Let us illustrate this for the second order term. In order not to make
the formulas too lengthy we will denote the propagator by:
$$
{\cal G}(x,y) = <x|G {1 \over 1+m G} |y>
$$
The first and second
derivative of the energy are now given by:
\be
{\partial E \over \partial m(x)} = {N \over 2} {\cal G} (x,x) +
2 N(N+2) g \int dy {\cal G}(y,y) {\cal G} (y,x) {\cal G} (x,y) + {\cal O} (g^2)
\ee
$$
{\partial^2 E \over \partial m(x)^2} = {\partial^2 E(g=0) \over m(x)^2} +
2 N(N+2) g \int dy  {\cal G} (y,x)^2 {\cal G} (x,y)^2 +
$$
\be
4 N(N+2) g \int dy  {\cal G} (y,x) {\cal G} (x,y)
{\cal G} (y,y) {\cal G} (x,x) + {\cal O} (g^2)
\ee
Again using this in (\ref{bootstrap}) we find after integration:
$$
E = {N \over 2} Tr \log(1 + m G) + N(N+2) g \int dx {\cal G}(x,x)^2 +
$$
$$
4 g^2 N(N+2) \int \int dx dy  {\cal G} (y,x)^2 {\cal G} (x,y)^2 +
$$
\be
4 g^2 N(N+2)^2 \int \int dx dy  {\cal G} (y,x) {\cal G} (x,y)
{\cal G} (y,y) {\cal G} (x,x) + {\cal O} (g^3)
\ee
Again by drawing the propagators or by considering the limit of constant
$m$ we find that this corresponds to the following diagramatic expression:

\begin{picture}(100,20)
%tweede orde bijdrage tot E
\put(10,10){$ 4 N(N+2)^2 \ g^2 \hspace{2.5 cm} + 4 N(N+2) \ g^2
\hspace{2 em} $}
\put(42,12){\circle{7}}
\put(49,12){\circle{7}}
\put(56,12){\circle{7}}
\put(100,12){\circle{8}}
\put(105,12){\circle{8}}
\end{picture}

which are indeed the correct diagrams with the correct symmetry factors.
It is clear that this procedure can be continued to any order one desires.
Of course the expressions will become rather elaborate as one advances
towards higher orders. For this reason we will present in the next
section a completely diagrammatic approach to the same problem, in which
one does not work with mathematical expressions, but with Feynman diagrams
that are combined together according to a set of simple rules.

\section{Diagrammatic approach}
The most important formula in developping this approach is formula
(\ref{double}) for which we have already given a graphical expression.
We will mainly repeat the steps followed in the preceding section, but
this time from a diagrammatic point of view. The lowest order
approximation of the energy is now given by:

\begin{picture}(100,20)
%E in laagste orde
\put(10,10){$E  = {N \over 2} \hspace{1.5cm} +{\cal O}(g)$}
\put(32,12){\circle{10}}
\end{picture}

\noindent
and the first functional derivative with respect to $m$(x) is given by (see
formula (\ref{der1})):

\begin{picture}(100,20)
%E in laagste orde eenmaal afgeleid
\put(10,10){${\partial E \over \partial m(x)} = {N \over 2} \hspace{1.5cm}
+ {\cal O}(g)$}
\put(38,12){\circle{10}}
\put(38,7){\circle*{1}}
\put(38,4){$x$}
\end{picture}

{}From now on we use formula (\ref{double}). In the above diagram we have
one propagator. Since the derivative doubles each propagator in turn
(and changes the sign) we find the following diagrammatic expression for
the second functional derivative:

\begin{picture}(100,20)
%E in laagste order tweemaal afgeleid
\put(10,10){${\partial^2 E \over \partial m(x)^2} = -{N \over 2}
\hspace{1.5cm} + {\cal O} (g)$}
\put(42,12){\circle{10}}
\put(42,7){\circle*{1}}
\put(42,17){\circle*{1}}
\put(42,4){$x$}
\put(42,19){$x$}
\end{picture}

Taking the integral with respect to $x$ in the diagrammatic approach
corresponds to identifying the different points that are labelled with "$x$"
and dropping the label. By doing this we create a vertex.
Doing so and using formula (\ref{bootstrap}) we find:

\begin{picture}(100,20)
%afgeleide van E naar g in laagste orde
\put(10,10){${\partial E \over \partial g} = N(N+2)$}
\put(45,12){\circle{7}}
\put(52,12){\circle{7}}
\put(48.5,12){\circle*{1}}
\end{picture}

\noindent
recovering the result found in the preceding section. We can now proceed
to higher orders by taking the functional derivative of the first order
approximation of the energy. This involves taking the derivative of the
the two loop diagram shown above. In this case we have two propagators,
so the derivative will be the sum of diagrams, in which every line is
doubled in turn (hereby changing the sign of the diagram).
Diagrammatically, this gives:

\begin{picture}(140,20)
%afleidingsregel voor eenvoudig diagram
\put(10,10){${\partial \over \partial m(x)}(\hspace{2cm}) = $}
\put(27,12){\circle{7}}
\put(34,12){\circle{7}}
\put(30.5,12){\circle*{1}}
\put(45,10){$= \ -$}
\put(62,12){\circle{7}}
\put(69,12){\circle{7}}
\put(58.5,12){\circle*{1}}
\put(54,10){$x$}
\put(75,10){$-$}
\put(86,12){\circle{7}}
\put(93,12){\circle{7}}
\put(96.5,12){\circle*{1}}
\put(98,10){$x$}
\put(104,10){$= - 2$}
\put(120,12){\circle{7}}
\put(127,12){\circle{7}}
\put(130.5,12){\circle*{1}}
\put(132,10){$x$}
\end{picture}

The second order derivative is taken in the same way. We now have
three propagators in each diagram, so doubling each of these in turn we
arrive at the expression:

\begin{picture}(100,20)
%tweede orde afgeleide van eenvoudig diagram
\put(10,10){${\partial^2 \over \partial m(x)^2}(\hspace{2cm}) = $}
\put(29,12){\circle{7}}
\put(36,12){\circle{7}}
\put(32.5,12){\circle*{1}}
\put(62,12){\circle{7}}
\put(69,12){\circle{7}}
\put(58.5,12){\circle*{1}}
\put(50,10){$2$}
\put(54,10){$x$}
\put(72.5,12){\circle*{1}}
\put(74,10){$x$}
\put(85,10){$+ 4$}
\put(96,12){\circle{7}}
\put(103,12){\circle{7}}
\put(103,15.5){\circle*{1}}
\put(103,8.5){\circle*{1}}
\put(103,17){$x$}
\put(103,5){$x$}
\end{picture}

\noindent
After integration with respect to $x$ this gives:

\begin{picture}(100,20)
%integraal van tweede afgeleide
\put(10,10){$\int \ dx {\partial^2 \over \partial m(x)^2}(\hspace{2cm}) = $}
\put(38,12){\circle{7}}
\put(45,12){\circle{7}}
\put(41.5,12){\circle*{1}}
\put(69,12){\circle{8}}
\put(74,12){\circle{8}}
\put(60,10){$2$}
\put(80,10){$+ 4$}
\put(91,12){\circle{7}}
\put(98,12){\circle{7}}
\put(105,12){\circle{7}}
\end{picture}

\noindent
Using the bootstrap formula (\ref{bootstrap}) once more we find the
diagrammatic equivalent of the expression found earlier in this paper:

\begin{picture}(100,40)
%E tot op derde orde
\put(40,30){${\partial E \over \partial g} = N(N+2)$}
\put(75,32){\circle{7}}
\put(82,32){\circle{7}}
\put(78.5,32){\circle*{1}}
\put(90,30){$+$}
\put(10,10){$ 8 N(N+2)^2 \ g \hspace{2.5 cm} + 8 N(N+2) \ g
\hspace{2 em} $}
\put(42,12){\circle{7}}
\put(49,12){\circle{7}}
\put(56,12){\circle{7}}
\put(100,12){\circle{8}}
\put(105,12){\circle{8}}
\put(115,10){$+ {\cal O}(g^2)$}
\end{picture}

It is clear how this can be continued to higher orders. Both taking the
functional derivative and integrating with respect to $x$ are translated
into simple diagrammatic manipulations, which allow a much simpler
treatment than the one presented in the preceding section. Carrying on
straightforwardly, one can easily generate Feynman diagrams to, in principle,
an arbitrary number of loops. We give here all diagrams up to five loops
with the correct symmetry factor for general N:

\begin{picture}(145,60)
\put(0,50){$E = {N \over 2} \hspace{.6 in} + N(N+2) g \hspace{.7 in} +
4 N(N+2)^2 g^2 \hspace{.8 in} + $}
\put(20,52){\circle{7}}
\put(60,52){\circle{7}}
\put(67,52){\circle{7}}
\put(105,52){\circle{6}}
\put(111,52){\circle{6}}
\put(117,52){\circle{6}}

\put(0,35){$4 N(N+2) g^2  \hspace{.6 in}  + 16 N(N+2)^3 g^3 \hspace{.8 in}
+64 N(N+2)^2 g^3 \hspace{.7 in}  +$}
\put(30,37){\circle{7}}
\put(34,37){\circle{7}}
\put(75,37){\circle{5}}
\put(80,37){\circle{5}}
\put(85,37){\circle{5}}
\put(90,37){\circle{5}}
\put(129,37){\circle{7}}
\put(133,37){\circle{7}}
\put(139,37){\circle{5}}

\put(0,20){${32 \over 3} N(N+2)(N+8) g^3 \hspace{.8 in} + {32 \over 3} N(N+2)^3
g^3
\hspace{.9 in}  + $}
\put(45,24){\circle{6}}
\put(51,24){\circle{6}}
\put(48,18.8){\circle{6}}
\put(97,20){\circle{6}}
\put(103,20){\circle{6}}
\put(109,20){\circle{6}}
\put(103,25.5){\circle{6}}

\put(0,5){$64 N(N+2)^4 g^4 \hspace{1.1 in} + 128 N(N+4)^4 g^4
\hspace{1 in} + $}
\put(33,7){\circle{5}}
\put(38,7){\circle{5}}
\put(43,7){\circle{5}}
\put(48,7){\circle{5}}
\put(53,7){\circle{5}}
\put(95,7){\circle{5}}
\put(100,7){\circle{5}}
\put(105,7){\circle{5}}
\put(110,7){\circle{5}}
\put(100,12){\circle{5}}
\end{picture}

\begin{picture}(125,60)
\put(0,50){$ 256 N(N+2)^3 g^4 \hspace{.9 in} + 256 N(N+2)^3 g^4
\hspace{.9 in} + $}
\put(33,52){\circle{5}}
\put(38,52){\circle{5}}
\put(44,52){\circle{7}}
\put(48,52){\circle{7}}
\put(92,52){\circle{7}}
\put(96,52){\circle{7}}
\put(101.2,49){\circle{5}}
\put(101.2,55){\circle{5}}

\put(0,35){$384 N(N+2)^3 g^4 \hspace{.9 in} + 256 N(N+2)^2 (N+8) g^4
\hspace{.9 in} + $}
\put(33,37){\circle{5}}
\put(39,37){\circle{7}}
\put(43,37){\circle{7}}
\put(49,37){\circle{5}}
\put(105,40){\circle{6}}
\put(105,34){\circle{6}}
\put(110.2,37){\circle{6}}
\put(115.7,37){\circle{5}}

\put(0,20){$ 32 N(N+2)^4 g^4 \hspace{.8 in} + 32 N(N+2)(N^2 + 6N +20) g^4
\hspace{.8 in} + $}
\put(31,22){\circle{6}}
\put(37,22){\circle{6}}
\put(43,22){\circle{6}}
\put(37,16){\circle{6}}
\put(37,28){\circle{6}}
\put(112,19){\circle{6}}
\put(112,24.5){\circle{6}}
\put(117.5,19){\circle{6}}
\put(117.5,24.5){\circle{6}}

\put(0,5){$ 64 N(N+2)(10 N+44) g^4 \hspace{.8 in} + 256 N(N+2)^2 g^4
\hspace{.8 in} $}
\put(55,3){\circle{6}}
\put(55,7){\circle{6}}
\put(55,5){\circle{10}}
\put(110,7){\circle{7}}
\put(114,7){\circle{7}}
\put(118,7){\circle{7}}
\end{picture}

\section{Correlation functions}
So far we have shown how one can generate the diagrams that contribute
to the energy of the model. If one were to use these diagrams to
calculate the perturbation series, one would inevitably be faced with
the problem of renormalisation. In order to be able to carry out the
renormalisation program, it is necessary to know the divergences of
the two- and four-point functions. Therefore the technique presented
in the preceding section would be of little use if it would not allow
the generation of these diagrams also. For this reason (and for the
sake of completeness) it is usefull to generalise the method to
the generation of any kind of diagram (vacuum, two-point, four-point,...).
As a matter of fact this can be easily accomplished in a straightforward
way by coupling a source term to the $\phi$-field. The model is then
defined by:
\be
{\cal Z} = e^{-E} = \int {\cal D} \phi \exp -{1 \over 2} \int
dx \partial_\mu \phi
\partial_\mu \phi + m(x) \phi ^2(x) - g \int (\phi ^2)^2 - \sum_{i=1}^N
\int J_i \ \phi_i \ ,
\ee
This extra term does
not break the validity of (\ref{bootstrap}), therefore the recursive
procedure can be used without any change. The only change we have to
deal with is the free field value of the energy. Instead of (\ref{free}) we
now find that $E(g=0)$ is given by:
\be
E(g=0) = {N \over 2} Tr \log(-\partial ^2 + m) -{1 \over 2}
\sum_{i=1}^N J_i.{1 \over (-\partial ^2 + m)}.J_i \ ,
\ee
where the "scalar product" in the last term is taken in coordinate space.
Since the diagrammatic approach is far simpler and neater we will use
it here from the very beginning. The above formula can be written
graphically as:

\begin{picture}(100,20)
%E met bron, in laagste orde
\put(10,10){$ E(g=0) = {N \over 2} \hspace{1cm} - {1 \over 2} J
\hspace{1.5 cm} J + {\cal O}(g)$}
\put(40,12){\circle{7}}
\put(57.5,10){\line(1,0){10}}
\end{picture}

The same rules for derivation and integration over a point in space
apply here. Taking the first and second derivative with respect to
$m(x)$ we find:

\begin{picture}(100,20)
%eerste afgeleide van E met bronterm in laagste orde
\put(10,10){${\partial E \over \partial m(x)} = {N \over 2} \hspace{1cm}
+ {1 \over 2} J \hspace{2 cm} J + {\cal O}(g)$}
\put(33,12){\circle{7}}
\put(33,8.5){\circle*{1}}
\put(33,5){$x$}
\put(50,10){\line(1,0){15}}
\put(58,10){\circle*{1}}
\put(58,5){$x$}
\end{picture}

\begin{picture}(100,20)
%tweede afgeleide van E met bronterm in laagste orde
\put(10,10){${\partial^2 E \over \partial m(x)^2} =- {N \over 2} \hspace{1cm}
- J \hspace{2 cm} J + {\cal O}(g)$}
\put(38,12){\circle{7}}
\put(38,8.5){\circle*{1}}
\put(38,15.5){\circle*{1}}
\put(38,5){$x$}
\put(38,17){$x$}
\put(53,10){\line(1,0){15}}
\put(57,10){\circle*{1}}
\put(57,5){$x$}
\put(63,10){\circle*{1}}
\put(63,5){$x$}
\end{picture}

\noindent
Putting this in (\ref{bootstrap}) and integrating with respect to $g$
we find the first order value of the energy:

\begin{picture}(140,20)
%E met bronterm in eerste orde
\put(10,10){$E = E(g=0) + ( N(N+2) \hspace{1.5cm} + (N+2) J \hspace{1.5 cm}
J + \hspace{1.5cm}) g +{\cal O}(g^2)$}
\put(65,12){\circle{5}}
\put(70,12){\circle{5}}
\put(98,10){\line(1,0){12}}
\put(104,12.5){\circle{5}}
\put(120,8){\line(1,1){8}}
\put(120,16){\line(1,-1){8}}
\put(117,8){$J$}
\put(117,16){$J$}
\put(129,8){$J$}
\put(129,16){$J$}
\end{picture}

It is quite clear how the recursion can be continued towards higher
orders. Doing this one generates $E$ as a functional of the source
$J$ to an arbitrary number of loops.
All correlation functions can then be obtained from this by taking
successive functional derivatives with respect to the source. As an
example we give here all one particle irreducible diagrams that
contribute to the two point function up to four loops:

\begin{picture}(120,100)
\put(10,75){${N+2 \over 2} g \hspace{1.1 cm} + {(N+2)^2 \over 2} g^2
\hspace{1.1 cm} + (N+2) g^2 $}
\put(25,78){\circle{6}}
\put(20,75){\line(1,0){10}}
\put(55,75){\circle{5}}
\put(55,80){\circle{5}}
\put(50,72.5){\line(1,0){10}}
\put(95,75){\circle{8}}
\put(89,75){\line(1,0){12}}

\put(10,60){${(N+2)^3 \over 2} g^3 \hspace{1.1cm} + (N+2)^2 g^3 \hspace{1.2cm}
+ 3(N+2)^2 g^3 $}
\put(30,60){\circle{5}}
\put(30,65){\circle{5}}
\put(30,70){\circle{5}}
\put(25,57.5){\line(1,0){10}}
\put(66,63){\circle{6}}
\put(66,66){\circle{6}}
\put(63,60){\line(1,0){6}}
\put(105,60){\circle{8}}
\put(105,66.5){\circle{5}}
\put(99,60){\line(1,0){12}}

\put(10,45){$(N+2)(N+8) g^3 \hspace{2.5cm} + {(N+2)^3 \over 2} g^3$}
\put(48,47){\circle{6}}
\put(54,47){\circle{6}}
\put(45,44){\line(1,0){12}}
\put(94,47){\circle{6}}
\put(90.1,50.9){\circle{5}}
\put(97.9,50.9){\circle{5}}
\put(91,44){\line(1,0){6}}

\put(10,30){${(N+2)^4 \over 2} g^4 \hspace{1.1 cm} + (N+2)^4 g^4
\hspace{1.5 cm} + {(N+2)^4 \over 2} g^4$}
\put(30,28){\circle{4}}
\put(30,32){\circle{4}}
\put(30,36){\circle{4}}
\put(30,40){\circle{4}}
\put(28,26){\line(1,0){4}}
\put(67,30){\circle{5}}
\put(63.82,33.18){\circle{4}}
\put(70.18,33.18){\circle{4}}
\put(73,36){\circle{4}}
\put(65,27.5){\line(1,0){4}}
\put(105,32){\circle{5}}
\put(105,37){\circle{5}}
\put(100,37){\circle{5}}
\put(110,37){\circle{5}}
\put(103,29.5){\line(1,0){4}}

\put(10,15){$(N+2)^3 g^4 \hspace{1.1cm} + 3(N+2)^3 g^4 \hspace{2 cm} +
2 (N+2)^3 g^4 $}
\put(35,15){\circle{4}}
\put(35,19.5){\circle{5}}
\put(35,22){\circle{5}}
\put(33,13){\line(1,0){4}}
\put(77,16){\circle{6}}
\put(77,21){\circle{4}}
\put(77,25){\circle{4}}
\put(72,16){\line(1,0){10}}
\put(118,17){\circle{5}}
\put(118,20.5){\circle{5}}
\put(122.5,17){\circle{4}}
\put(115,14.5){\line(1,0){6}}
\end{picture}

\begin{picture}(120,75)
\put(10,75){$3(N+2)^3 g^4 \hspace{2cm} + 3(N+2)^3 g^4 \hspace{1.1 cm}
+3 (N+2)^3 g^4 $}
\put(40,76){\circle{6}}
\put(43.54,79.54){\circle{4}}
\put(36.46,79.54){\circle{4}}
\put(35,76){\line(1,0){10}}
\put(85,77){\circle{5}}
\put(85,80.5){\circle{5}}
\put(85,85){\circle{4}}
\put(83,74.5){\line(1,0){4}}
\put(125,77){\circle{6}}
\put(125,82){\circle{4}}
\put(125,72){\circle{4}}
\put(120,77){\line(1,0){10}}

\put(10,60){$4(N+2)^2 (N+8) g^4 \hspace{2 cm} + (N+2)^2 (N+8) g^4 $}
\put(50,62){\circle{5}}
\put(55,62){\circle{5}}
\put(58.2,65.2){\circle{4}}
\put(48,59.5){\line(1,0){10}}
\put(115,60){\circle{5}}
\put(117.5,64.33){\circle{5}}
\put(112.5,64.33){\circle{5}}
\put(113,57.5){\line(1,0){4}}

\put(10,45){$(N+2)^2 (N+8) g^4 \hspace{2cm} + {(N+2)^4 \over 2} g^4$}
\put(50,47){\circle{5}}
\put(55,47){\circle{5}}
\put(48,44.5){\line(1,0){10}}
\put(52.5,42.5){\circle{4}}
\put(95,46){\circle{6}}
\put(100,46){\circle{4}}
\put(90,46){\circle{4}}
\put(95,51){\circle{4}}
\put(93,43){\line(1,0){4}}

\put(10,30){$(N+2)(N^2+6 N+20) g^4 \hspace{2.5 cm} +
(N+2)(10N+44) g^4$}
\put(63,32){\circle{5}}
\put(71,32){\circle{5}}
\put(67,33){\circle{4}}
\put(61,29.5){\line(1,0){12}}
\put(132,30){\circle{8}}
\put(132,32){\circle{4}}
\put(126,30){\line(1,0){12}}

\put(10,15){$(N+2)(10N+44) g^4 \hspace{2.5 cm}+ 6(N+2)^2 g^4$}
\put(60,14){\circle{8}}
\put(53.5,14){\circle{5}}
\put(51,11.5){\line(1,0){15}}
\put(110,15){\circle{8}}
\put(110,19){\circle{4}}
\put(104,15){\line(1,0){12}}
\end{picture}

\section{Gross Neveu model}
In this section we will present an analogous treatment of the Gross-Neveu
model \cite{grne}. This well known model is defined by the partition function:
\be
\label{GN}
{\cal Z} = e^{-E} = \int {\cal D} \overline{\psi} {\cal D} \psi
e^{- \int \overline{\psi} {\not \partial} \psi  - g (\overline{\psi} \psi)^2}
\ee
In order to obtain the perturbation series for this massless model we
will consider the case where the field is allowed to have a mass m, which
again depends on the coordinates. Instead of working with (\ref{GN}), we
will investigate the theory with partition function:
\be
{\cal Z} = e^{-E} = \int{\cal D} \overline{\psi} {\cal D} \psi
e^{- \int \overline{\psi} ({\not \partial}+ m(x)) \psi - g (\overline{\psi}
\psi)^2}
\ee
Once again we find a relation between the partial (functional) derivatives
of the energy:
\be
\label{bootstrap2}
{\partial E \over \partial g} = \int (({{\partial E} \over
{\partial m(x)}}) ^2  - {{\partial ^2 E} \over {\partial m(x)} ^2})
\ee
The only missing ingredient is the value of the energy at zero coupling.
Again without difficulty one finds that this is given by:
\be
E(g=0)=-N Tr \log({\not \partial}+m) = -N Tr \log(1+ m {\not \partial}^{-1}) +
cte
\ee
In this case "$Tr$" means the trace taken in coordinate space as well
as the trace of the $\gamma$-matrices that occur in the expression (one
could write it as: $Tr = Tr^{(\gamma)} \otimes Tr^{(coord)}$, where
$Tr^{(\gamma)}$ denotes the trace of $\gamma$-matrices and $Tr^{(coord)}$
is the trace in coordinate space).
This is an important difference with the first model considered
in this paper. We can also derive the
equivalent formulae of (\ref{der1}) and (\ref{double}), namely:
\be
{\partial E \over m(x)} = - N Tr^{(\gamma)} <x|G {1 \over 1+m G} |x>
\ee
\noindent
and:
\be
{\partial  \over \partial m(x)} <y|G {1 \over 1+ m G}|z> =
-<y| G {1 \over 1+ m G} |x> <x| G {1 \over 1+ m G} |z>
\ee
The last formula is in fact the same as (\ref{double}), except for the
fact that G is now ${\not \partial}^{-1}$. \newline
Instead of working with the rather cumbersome mathematical expressions,
we will immediately turn to the diagrammatic approach proposed in the
preceding section (with a few minor changes to make it better adapted
to this model).
The lowest order approximation of the energy could be written as:

\begin{picture}(100,20)
%E van GN-model
\put(10,10){$ E = - N \hspace{1 cm} + {\cal O}(g)$}
\put(31,12){\circle{7}}
\end{picture}

It is however more convenient to attribute a minus sign and a factor of
$N$ to the closed loop itself, so that the energy in diagrammatic notation
would simply be:

\begin{picture}(100,20)
%E van GN-model met N opgeslorpt in de lus
\put(10,10){$E = \hspace{1cm} + {\cal O}(g)$}
\put(23,12){\circle{7}}
\end{picture}

We now proceed to generate the diagrams with this "initial value" and
remember to assign to a certain diagram a power of $N$ equal to the
number of loops in the diagram and a minus sign if this number is odd.
Moreover, around every loop one would have to take a trace of the
$\gamma$-matrices in the loop. Let us make this clear in the lowest
orders. The first and second functional derivative are given by:

\begin{picture}(100,20)
%E van GN-model, eerste afgeleide
\put(10,10){$ {\partial E \over \partial m(x)} = \hspace{1 cm} + {\cal O}(g)$}
\put(30,12){\circle{7}}
\put(30,8.5){\circle*{1}}
\put(30,5){$x$}
\end{picture}

\begin{picture}(100,20)
%E van GN-model, tweede afgeleide
\put(10,10){$ {\partial^2 E \over \partial m(x)^2} = - \hspace{2 cm}
+ {\cal O}(g)$}
\put(40,12){\circle{7}}
\put(36.5,12){\circle*{1}}
\put(43.5,12){\circle*{1}}
\put(33,10){$x$}
\put(45,10){$x$}
\end{picture}

We still have the property that taking
the functional derivative doubles a propagator
in the diagram and changes the sign. The main difference is that the
propagator is now the propagator of a fermion field. Using the bootstrap
formula, we find for the energy of the Gross-Neveu model in first
order:

\begin{picture}(100,20)
%E van GN model in eerste orde
\put(10,10){$E = \hspace{1 cm} + (\hspace{3 cm})g + {\cal O}(g^2)$}
\put(23,12){\circle{7}}
\put(37,12){\circle{5}}
\put(47,12){\circle{5}}
\put(51,10){$+$}
\put(60,12){\circle{7}}
\multiput(39.5,12)(1,0){5}{$.$}
\multiput(56.5,12)(1,0){7}{$.$}
\end{picture}

In the last formula we have slightly changed the convention used in the
preceding section. Instead of merging two points with the same label,
we now connect them with a dotted line, reminiscent of the auxiliary
field notation. We do this in order to be able to keep track of the
number of (index-)loops in each diagram, which is important for the "weight"
of factors of $N$ that the diagram will receive,
as well as for the sign of the diagram. If one thinks of the
dotted line as a delta-function it is easy to see that this is indeed
the same as the convention in the preceding section. The next order in
the coupling constant is found in much the same way:

\begin{picture}(100,20)
%eerste afgeleide van E(GN), eerste orde
\put(10,10){${\partial E \over \partial m(x)} =
\hspace{1 cm} - 2 g \hspace{2.3cm} - 2 g \hspace{1.5 cm} + {\cal O}(g^2)$}
\put(30,12){\circle{7}}
\put(47,12){\circle{5}}
\put(57,12){\circle{5}}
\multiput(49.5,12)(1,0){5}{$.$}
\put(80,12){\circle{7}}
\multiput(76.5,12)(1,0){7}{$.$}
\put(59.5,12){\circle*{1}}
\put(61,10){$x$}
\put(80,8.5){\circle*{1}}
\put(80,5){$x$}
\put(30,8.5){\circle*{1}}
\put(30,5){$x$}
\end{picture}

\noindent
and:

\begin{picture}(100,40)
%tweede afgeleide van E(GN), eerste orde
\put(10,30){${\partial^2 E \over \partial m(x)^2} = -
\hspace{1 cm} + 2 g \hspace{2.5cm} +4 g \hspace{2.5 cm} +  $}
\put(35,32){\circle{7}}
\put(35,28.5){\circle*{1}}
\put(35,25){$x$}
\put(35,35.5){\circle*{1}}
\put(35,35.5){$x$}
\put(57,32){\circle{5}}
\put(67,32){\circle{5}}
\multiput(59.5,32)(1,0){5}{$.$}
\put(69.5,32){\circle*{1}}
\put(71,30){$x$}
\put(54.5,32){\circle*{1}}
\put(51,30){$x$}
\put(87,32){\circle{5}}
\put(97,32){\circle{5}}
\multiput(89.5,32)(1,0){5}{$.$}
\put(97,29.5){\circle*{1}}
\put(97,26){$x$}
\put(97,34.5){\circle*{1}}
\put(97,36){$x$}
\put(30,10){$2 g \hspace{1.5cm} + 4 g \hspace{2cm} + {\cal O} (g^2)$}
\put(42,12){\circle{7}}
\multiput(38.5,12)(1,0){7}{$.$}
\put(42,8.5){\circle*{1}}
\put(42,5){$x$}
\put(42,15.5){\circle*{1}}
\put(42,15.5){$x$}
\put(64,12){\circle{7}}
\multiput(60.5,12)(1,0){7}{$.$}
\put(62.25,9){\circle*{1}}
\put(65.75,9){\circle*{1}}
\put(61,5){$x$}
\put(66.5,5){$x$}
\end{picture}

\noindent
which leads to the energy approximated in second order:

\begin{picture}(100,40)
%tweede orde bijdrage tot E(GN)
\put(10,30){$8 g^2 \hspace{3 cm} + 16 g^2 \hspace{2.1 cm} +
4 g^2 $}
\put(30,10){$ 4 g^2 \hspace{1.5cm} + 4 g^2 \hspace{1.5 cm}$}
\put(20,32){\circle{5}}
\put(30,32){\circle{5}}
\put(40,32){\circle{5}}
\put(63,32){\circle{7}}
\put(74,32){\circle{5}}
\put(95,32){\circle{6}}
\put(107,32){\circle{6}}
\put(43,12){\circle{8}}
\put(68,12){\circle{8}}
\multiput(22.5,32)(1,0){5}{$.$}
\multiput(32.5,32)(1,0){5}{$.$}
\multiput(63,28.5)(0,1){7}{$.$}
\multiput(66.5,32)(1,0){5}{$.$}
\multiput(97.6,30.4)(1,0){7}{$.$}
\multiput(97.6,33.6)(1,0){7}{$.$}
\multiput(39,12)(1,0){8}{$.$}
\multiput(43,8)(0,1){8}{$.$}
\multiput(65.17,14.83)(0,-1){7}{$.$}
\multiput(69.83,14.83)(0,-1){7}{$.$}
\end{picture}

The notation with the "auxiliary field" is clearly more convenient
when it comes to taking traces of $\gamma$-matrices. The more conventional
diagrams would obscure the way these traces have to be taken.
When one wants to consider the massless case (as the Gross-Neveu
model is originally defined), a number of diagrams will automaticly
yield zero. Indeed, in this case the propagator is given by ${{\not \partial}
\over \partial^2}$
and therefore every propagator contributes one $\gamma$-matrix. This
means that loops with an odd number of propagators automatically vanish.
(this includes for instance all tadpole diagrams).
Note that one cannot discard these diagrams before one has arrived
at the end of the calculation. To obtain higher order diagrams correctly,
one has to take into account all diagrams of lower order, also those that
would vanish in the limit when $m$ is taken to zero.
Correlation functions for the GN-model can likewise be generated by
adding a source term. In the fermionic case however, one would have to
introduce two source terms:
\be
{\cal Z} = e^{-E} = \int{\cal D} \overline{\psi} {\cal D} \psi
e^{-\int \overline{\psi} ({\not \partial}+ m(x)) \psi - g (\overline{\psi}
\psi)^2
-\overline{\eta} \psi -\overline{\psi} \eta }
\ee
The lowest order approximation to the energy would then read:

\begin{picture}(100,20)
\put(10,10){$E =  \hspace{1.5 cm} - \overline{\eta} \hspace{1.5 cm} \eta$}
\put(28,12){\circle{7}}
\put(42,10){\line(1,0){10}}
\end{picture}

\noindent
and one could use (\ref{bootstrap2}) to generate the energy as a
functional of $\eta$ and $\overline{\eta}$. Correlation functions are
easily found by taking successive derivations with respect to the
source terms.

\section{Thirring model}
The third model which can be treated in a way similar to the vector model
and the Gross-Neveu model is the Thirring model \cite{thir}.
This model also is a theory
of interacting fermions, but this time with an interaction term
$-g \sum_{\mu} \int (\overline{\psi} \gamma_{\mu} \psi)^2$.
In order to be able to carry
out the same techniques as for the previous model, it is not sufficient to
consider an $x$-dependent mass. In this case we must take a mass that also
depends on the Dirac indices. We will define this "matrix-mass" as:
${\not m} = \gamma ^{\mu} m_{\mu} (x)$. The generalised partition function
that we consider is now given by:
\be
{\cal Z} = e^{-E} = \int {\cal D} \overline{\psi} {\cal D} \psi e^{
-\int \overline{\psi}({\not \partial} +{\not m}) \psi -
g (\overline{\psi} \gamma_{\mu} \psi)^2}
\ee
The formulas for the functional partial derivatives are given by:
$$
{{\partial E} \over {\partial m_{\mu}(x)}} = <\overline{\psi} \gamma_{\mu}
\psi(x)>
$$
\noindent
and:
\be
{{\partial ^2 E} \over {\partial m_{\mu}(x)} ^2} =
-<(\overline{\psi} \gamma_{\mu} \psi(x))^2>  +
({{\partial E} \over {\partial m_{\mu}(x)}}) ^2,
\ee
The bootstrap formula is easily generalised to:
\be
\label{bootstrap3}
{\partial E \over \partial g} = \sum _{\mu} \int dx (({{\partial E} \over
{\partial m_{\mu}(x)}}) ^2  - {{\partial ^2 E} \over {\partial m_{\mu}(x)} ^2})
\ee
We also need to know the value of the energy when the interaction is
switched of. Once more, this poses no serious problems and we find:
\be
E=-N Tr \log({\not \partial}+{\not m}) = -N Tr \log(1+ {\not m} {\not
\partial}^{-1}) + cte
\ee
Just as in the Gross-Neveu model the trace is to be understood as a
direct product of the trace in coordinate space and the trace of the
$\gamma$-matrices in the expression. The formula for the functional
derivative of an propagator is slightly different from the forms
encountered above. We find without problems:
\be
{\partial E \over \partial m_{\mu}(x)} <y|G {1 \over 1+ m G}|z> =
-<y| G {1 \over 1+ m G} |x> \gamma _{\mu} <x| G {1 \over 1+ m G} |z>
\ee
This means that, apart from a doubling of the propagator and a changing
of sign, the derivative inserts a $\gamma$-matrix at the point $x$. This
poses no serious difficulties however in the diagrammatic expansion.
As a matter of fact, one obtains the same diagrams as in the Gross-
Neveu model (again using the notation with dotted lines to connect
identical points). The only difference is the value that has to be assigned
to each of these diagrams. The trace of $\gamma$-matrices around
a loop now contains not only the matrices from the propagators in the
loop, but also those inserted at the "vertices" of the fermion field
and the "auxiliary field". Because of this, the number of $\gamma$-
matrices in a loop is always even. Therefore, unlike the massless Gross-Neveu
model, all diagrams will contribute.
The problem of generating diagrams that contribute to the various
correlation functions can be treated by the addition of two extra
source terms as in the case of the GN model.

\section{Non-local interactions - QED}
Another interesting generalisation of the method presented in this paper
is the case of non-local interactions. Instead of an interaction term
$\int \ dx (\phi^2(x))^2$ (or the analogon for the other models considered
here) one could work with an interaction term:
\be
\int \ dx \; dy \phi^2 (x) f(x,y) \phi^2 (y)
\ee
\noindent
where $f(x,y)$ is some function of $x$ and $y$. The diagrams that
contribute to the energy of this model are not very much different
from the diagrams of the models we have considered in the preceding
sections. As a matter of fact, when we use the auxiliary field notation they
are exactly the same! The only difference is that up to now the
propagator of the auxiliary field was merely regarded as a notation
for the $\delta$-function and that in the end one would have to identify
the points that are connected by such a propagator. In the case of non-
local interactions the propagator is a non-local function and the
dotted lines now denote a real field propagating. This becomes clear
when one considers the bootstrap equation for non-local interactions.
We easily find:
\be
\label{bootstrapnl}
{\partial E \over \partial g} = 4 \int \ dx \; dy f(x,y) (({{\partial E} \over
{\partial m(x)}}) ({\partial E \over \partial m(y)})
- {{\partial ^2 E} \over \partial m(x) \partial m(y)})
\ee
If one uses the diagrammatic approach, one easily finds that the
recursive method now comes down to marking two {\em different} points
($x$ and $y$) in the diagrams and connecting them with a line which,
in this case, represents a real field propagating between the two
points. The functions $f(x,y)$ can be seen as the propagator of the
field in question.

The most important example of a theory that has this
kind of interaction is of course QED. QED is defined by the partition
function:
\be
Z= e^{-E}= \int {\cal D} A {\cal D} \psi {\cal} \overline{\psi}
e^{-\int {1 \over 2} \partial A_{\mu} . \partial A^{\mu}
+ \overline{\psi} ( \not \partial + m + g \gamma_{\mu} A^{\mu}) \psi }
\ee
\noindent
Since the action is quadratic in the photon field we can easily
integrate this out and we end up with:
\be
Z= e^{-E}= \int {\cal D} \psi {\cal} \overline{\psi}
e^{ - \int \overline{\psi} ( \not \partial + m) \psi + \int \; dx \; dy
\overline{\psi} \gamma_{\mu} \psi (x) \Delta^{\mu \nu} (x-y)
\overline{\psi} \gamma_{\nu} \psi (x)} \ ,
\ee
where $\Delta$ is nothing but the photon propagator. This action is
just a Thirring model with non-local interaction. Therefore the technique
we have presented also allows one to generate Feynman diagrams that
contribute to the energy and the various correlation functions of QED.

\section{Conclusion and outlook}
We have presented in this paper a method that allows to generate Feynman
diagrams for theories that involve fields with $N$ components, both
bosonic and fermionic. By applying a simple algoritm one obtains
diagrams of a certain order from the knowledge of the diagrams of lower
order. A simple, purely diagrammatic, method for this can be constructed.
In this paper we have restricted ourselves to quartic interactions, but
one clearly sees that interactions of higher order can also be treated
in a similar way. In this case the equations (\ref{bootstrap}),
(\ref{bootstrap2}) and
(\ref{bootstrap3}) will contain higher order derivatives with respect to
$m(x)$. For a interaction of order $2 n$ one would have derivatives up
to $n^{th}$ order. One is however always restricted to even order
interaction terms (which was to be expected, since these are the only ones
that make sense for general vector models with $O(N)$-symmetry).

The techniques presented in this paper apply only to vector models so far.
It would of course be interesting to have a similar technique for the
matrix models \cite{sig}. Other interesting extensions are the
generation of diagrams
in theories with odd interaction terms and theories with several interaction
terms. The most important example of such a theory is of course QCD.

\end{document}